\documentclass[12pt]{article}
\usepackage{epsfig,amsfonts,amssymb}
\usepackage{cite,epsf}

\newcommand{\be}{\begin{equation}}
\newcommand{\ee}{\end{equation}}
\newcommand{\ben}{\begin{eqnarray}\displaystyle}
\newcommand{\een}{\end{eqnarray}}

\newcommand{\bea}[1]{\begin{eqnarray}\label{#1} }
\newcommand{\eea}{\end{eqnarray}}

\newcommand{\refb}[1]{(\ref{#1})}

\newcommand{\sectiono}[1]{\section{#1}\setcounter{equation}{0}}

\newcommand{\ket}[1]{\left| #1\right\rangle}

\def\one{{\hbox{ 1\kern-.8mm l}}}
\def\zero{{\hbox{ 0\kern-1.5mm 0}}}

\begin{document}
\begin{titlepage}

\title{
{\Large\bf Contour Integral Representations for the Characters of
Logarithmic CFTs}}

\author{
{}\\
{\large\bf Arjun Bagchi, Turbasu Biswas}\\
{\large\it Harish-Chandra Research Institute}\\
{\large\it Chhatnag Road, Jhusi}\\
{\large\it Allahabad 211019, India}\\
{\tt arjun, turbasu@hri.res.in}\\
{}\\
and\\
{}\\
{\large\bf Debashis Ghoshal}\thanks{On leave of absence from
the Harish-Chandra Research Institute, Allahabad, India.}\\
{\large\it School of Physical Sciences}\\
{\large\it Jawaharlal Nehru University}\\
{\large\it New Delhi 110067, India}\\
{\tt dghoshal@mail.jnu.ac.in}\\
{}\\
}

\bigskip\bigskip

\date{
\begin{quote}
\centerline{\bf Abstract:} We propose a contour integral
representation for the one-point correlators at genus one of the
primaries of a family of rational logarithmic conformal field
theories.
\end{quote}
\bigskip\bigskip
\leftline{{\bf Report No: HRI-P-08-10-001}} }


\end{titlepage}
\maketitle\vfill \eject

\tableofcontents

\sectiono{Introduction}\label{introd}
Conformally invariant quantum field theories in two dimensions have
been widely studied because of their relevance to diverse phenomena.
Of considerable interest is the classification of conformal field
theories (CFTs). There has been significant progress in classifying
a subset of CFTs, namely the rational conformal field theories
(RCFTs). These are theories in which the number of primary fields
(with respect to the infinite dimensional Virasoro algebra or an
extended chiral algebra) is finite.

Among the diverse approaches to the problem, the idea in
Refs.\cite{Eguchi:1987qd}--\cite{Eguchi:1988wh} to consider the
one-loop characters of the irreducible representations of the chiral
algebra is a beautiful one. They propose to view these as the
independent solutions of a differential equation of finite order.
Since these characters transform into linear combinations of
themselves under a modular
transformation\cite{Cardy:1986ie,Nahm:1991ie}, the differential
equation itself must have definite covariance property under a
modular transformation. CFTs with two or three characters have been
classified in this approach. The differential equation has been
known to be related to the existence of a null
vector\cite{Eguchi:1987qd} (see also
\cite{Gaberdiel:2008pr,Gaberdiel:2008ma} for a more recent and
rigorous analysis).

CFTs have since been generalised to accommodate {\em logarithmic
conformal field theories} (LCFTs). A characteristic feature of the
theories in this class is the existence of a logarithmic branch cut
in certain chiral correlation functions\cite{Gurarie:1993xq}. The
theory studied in \cite{Gurarie:1993xq} is the first member of a
series of perhaps the simplest LCFTs, the logarithmic $(p,1)$ {\em
minimal models}. These theories are rational with respect to a
chiral $W$-algebra\cite{Kausch:1990vg,Gaberdiel:1996np}. However,
not all the highest weight representations are completely
decomposable\cite{Rohsiepe:1996qj}. LCFTs have been studied
extensively since their discovery (a partial list is
Refs.\cite{Kausch:1995py}--\cite{Flohr:2007jy}). There are many
examples where the nature of representations differ from the minimal
family that we will focus on. For recent reviews, see, for example,
Refs.\cite{Flohr:2001zs, Gaberdiel:2001tr}.

In this paper, we consider modular differential equation for the
logarithmic $(p,1)$ minimal models. These have been studied in
Ref.\cite{Flohr:2005cm}, where it is shown that the vacuum torus
amplitudes can be thought to arise as solutions to a modular
differential equation. Our focus, however, will be on finding
explicit integral representations for the vacuum torus amplitudes
along the lines of Ref.\cite{Mukhi:1989qk}. The advantage of the
integral representations is that they are explicit, one can in
principle calculate the solution to any give order. Moreover, it is
shown in the recent paper \cite{Gaberdiel:2008ma} that while the
torus one-point functions in an RCFT satisfy a modular differential
equation, the solutions are not always expressible in terms of
standard transcendental functions. In view of this result the
integral representations could be an effective way to find these
amplitudes. They may also be useful in the calculation of the other
correlation functions on the torus.

In the following, we review the logarithmic correlator of
Ref.\cite{Gurarie:1993xq} in Sec.\ref{lmm} and write the integral
representations of the hypergeometric differential equation. We also
recall some relevant facts about the LCFTs. In Sec.\ref{mde}, we
review the classification of RCFTs in terms of the modular
differential equation. We also comment on the case for LCFTs. In
Secs.\ref{contourcm2} and \ref{contourlmm}, the integral
representations for the characters, more precisely, the vacuum torus
amplitudes is proposed. We consider this in some detail for the
simplest case of the $(2,1)$ minimal model with $c=-2$
(Sec.\ref{contourcm2}) and comment on the more general models
(Sec.\ref{contourlmm}). We end with some concluding remarks.

\sectiono{Logarithmic minimal models}\label{lmm}
An infinite family of logarithmic CFTs is the set of minimal models
labelled by $(p,1)$ with $p=2,3,\cdots$ following the notation of
BPZ\cite{Belavin:1984vu}. The central charge of the $(p,1)$ model
${\cal M}_p$ is
\begin{equation}
c_p = 1 - {6(p-1)^2\over p} = 13 - \left(p + {1\over p}\right)
\label{ccharge}
\end{equation}
and the degenerate fields $\phi_{r,s}$, $r,s\in\mbox{\bf Z}^+$ have
the conformal dimensions
\begin{equation}
h_{r,s} = {(sp-r)^2 - (p-1)^2\over 4p}\label{cweights}
\end{equation}
and a null vectors at level $rs$. The set of primaries inside the
Kac table for these models is empty. There is, however, a chiral
$W$-algebra generated by fields of spin $(2p-1)$. Restricting the
set of degenerate primaries to the range $s=1$ and $1\le r\le 3p-1$
yields a finite dimensional representation\cite{Flohr:2001zs}. We
may think of these fields as being on the boundary of an extended
Kac table.

The first of these models corresponding to $p=2$ has $c=-2$. The
set of primaries $\phi_{r,1}\equiv \phi_r$ with $r=1,\cdots,5$ are
in representations of the chiral $W_3$-algebra generated by a
triplet of spin-3 fields\cite{Kausch:1990vg,Gaberdiel:1996np}:
The fields $\phi_1$, $\phi_2$ and $\phi_3$ with weights $0$, $-1/8$
and $0$ respectively are singlet
representations, and $\phi_4$ and $\phi_5$ with weights $3/8$ and
$1$ are doublets. Moreover, while the doublets as well as the singlet
$\phi_2$ are irreducible representations of the Virasoro algebra,
the pair $\phi_1$ and $\phi_3$ (with $h_1=h_3$) form an indecomposable
Jordan block. For example, under the action of $L_0$:
\begin{eqnarray}
L_0\,\ket{\phi_1} &=& h_1\ket{\phi_1},\nonumber\\
L_0\,\ket{\phi_3} &=& h_3\ket{\phi_3} + \ket{\phi_1}.\label{jpair}
\end{eqnarray}
This is a characteristic feature of these logarithmic theories. In
the model ${\cal M}_p$, there are $(p-1)$ pairs of equal weights
that form $2\times 2$ indecomposable Jordan blocks of the Virasoro
algebra. These are the fields $(\phi_1,\phi_{2p-1}), (\phi_2,
\phi_{2p-2}),\cdots,(\phi_{p-1},\phi_{p+1})$, all (except the first)
of which have negative weights\footnote{Indeed as argued in
Ref.\cite{Cardy:1999zp} the logarithmic dependence can arise only in
non-unitary theories.} $h\le 0$ (equality for the first). The
remaining fields $\phi_p$ and $\phi_{2p},\cdots,\phi_{3p-1}$ are all
in irreducible representations, with all except the first field
having positive weights.

Coming back to the $c=-2$ model ${\cal M}_2$, the field $\phi_2$
has a null vector at level two. This leads to a differential
equation for the four-point correlation function\cite{Gurarie:1993xq}
\begin{equation}
\Big\langle \phi_2(z_1)\phi_2(z_2)\phi_2(z_3)\phi_2(z_4)
\Big\rangle \sim \left[(z_1-z_2)(z_3-z_4)\right]^{1/4}\,
\left[\xi(1-\xi)\right]^{1/4}F(\xi)\label{fourspin}
\end{equation}
with
\begin{equation}
\xi(1-\xi){d^2F\over d\xi^2} + (1-2\xi){dF\over d\xi} - {1\over4}
F = 0, \label{hyperg}
\end{equation}
where, $\xi$ is the cross-ratio and we have only displayed the
chiral part of the correlator. This is a hypergeometric differential
equation with $a=b=1/2$ and $c=1$. The two solutions
\begin{eqnarray}
I_1 &=& F(a,b,c;\xi) = {\Gamma(c)\over\Gamma(b)\Gamma(c-b)}
\int_1^\infty dt\,t^{a-c}(1-t)^{c-b-1}(t-\xi)^{-a},\nonumber\\
I_2 &=& \xi^{1-c} F(b-c+1,a-c+1,2-c;\xi)\nonumber\\
&=& {\Gamma(2-c)\over\Gamma(1-a)\Gamma(1+a-c)}
\int_0^\xi dt\,t^{a-c}(1-t)^{c-b-1}(\xi-t)^{-a},
\label{hgIsolns}
\end{eqnarray}
for this choice of parameters are not independent, and indeed can be
obtained from one another by a change of integration variable. The
second solution, therefore, is to be replaced by one with a
logarithmic dependence. The Jordan block structure of the two $h=0$
fields of the theory was deduced from it in
Ref.\cite{Gurarie:1993xq}.

In order to obtain the logarithmic piece, we may replace the
parameters $a$, $b$ and $c$ by adding an infinitesimal piece
$\varepsilon$ to them (one can do this for $c$ only, but this is
more general). Now consider the linearly independent combinations
$\tilde{I}_1\equiv\left(I_1+I_2\right)/2=I_1$ and
$\tilde{I}_2\equiv\displaystyle\lim_{\varepsilon\to 0}
\left(I_1-I_2\right)/\varepsilon\sim \ln\xi\,I_1(\xi) + P(\xi)$,
where $P(\xi)$ is an infinite series in $\xi$ that can be
determined. A similar regularisation to obtain the logarithmic
solution was done in Ref.\cite{Hata:2000zg}.

\sectiono{The modular differential equation}\label{mde}
In a rational conformal field theory, the number of primaries with
respect to the conformal Virasoro algebra or a larger chiral algebra
is finite. As a result, the partition function as well as the correlation
functions on an arbitrary genus Riemann surface, may be expressed as a
sum of products of a finite number of holomorphic and anti-holomorphic
building blocks. These are functions of the various moduli. For the one
loop partition function, these holomorphic functions are also the
(appropriately defined) characters of the representations of the symmetry
algebra\cite{Cardy:1986ie}. Under a modular transformation of the torus,
the characters transform as linear combinations of themselves and thus
provide a representation of the modular group of the genus-one surface.

The authors of Refs.\cite{Eguchi:1987qd}--\cite{Mathur:1988gt} proposed
to classify RCFTs in terms of their characters. The finite number $n$ of
the characters are to be regarded as the independent solutions of a
differential equation of order $n$. Given the transformation properties
of the characters, this differential equation must be modular invariant.
In order to write this equation, one must take note of the fact that the
derivative $\partial_\tau\equiv{\partial\over\partial\tau}$ does not
transform covariantly under a modular transformation. Instead, one
applies the covariant derivative on a modular form of degree $2k$
\begin {equation}
{\mathcal{D}}_{(k)} = \partial_\tau - {i\pi k\over6} E_2(\tau),
\label{modcovd}
\end{equation}
(where $E_2(\tau)$ is the second Eisenstein series), appropriately to
write the most general modular invariant differential equation (MDE)
of order $n$:
\begin{equation}
{\mathcal{D}}_{\tau}^n\chi + \sum_{k=0}^{n-1} f_k(\tau)
{\mathcal{D}}_{\tau}^k\chi = 0.\label{moddiffe}
\end{equation}
In the above $f_k(\tau)$ are modular forms of weight $2(n-k)$. Let
$\chi_1(\tau),\cdots,\chi_n(\tau)$ be $n$ linearly independent solutions
of the MDE. The coefficients are then
$f_k(\tau) = (-)^{n-k}W_k(\tau)/W(\tau)$, where
\begin{equation}
W_k(\tau) = \det \pmatrix{\chi_1 & \chi_2 & \cdots & \chi_n\cr
\mathcal{D}_\tau\chi_1 & \mathcal{D}_\tau\chi_2 & \cdots &
\mathcal{D}_\tau\chi_n\cr
\vdots & \vdots & \ddots & \vdots\cr
{\mathcal{D}}_\tau^{k-1}\chi_1 & \mathcal{D}_\tau^{k-1}\chi_2 & \cdots
& {\mathcal{D}}_\tau^{k-1}\chi_n\cr
{\mathcal{D}}_\tau^{k+1}\chi_1 & \mathcal{D}_\tau\chi_2 & \cdots
& {\mathcal{D}}_\tau^{k+1}\chi_n\cr
\vdots & \vdots & \ddots & \vdots\cr
{\mathcal{D}}_\tau^n\chi_1 & \mathcal{D}_\tau^n\chi_2 & \cdots
& {\mathcal{D}}_\tau^n\chi_n}
\label{wronsk}
\end{equation}
and $W(\tau)\equiv W_n(\tau)$ is the Wronskian.

As explained in \cite{Mathur:1988na,Mathur:1988gt}, the classification of
RCFTs is then characterised by two numbers: $n$, the number of characters
or the order of the MDE and $\ell$, the number of zeros\footnote{Due to the
presence of the orbifold points in the moduli space, $6\ell$, and not $\ell$
is required to be an integer.} of the Wronskian $W$. The number $\ell$ can
in turn be expressed in terms of $n$, the central charge $c$ and the
weights $h_\alpha$ of the primary fields as
$\ell={1\over2}n(n-1)+{1\over4}nc-6\sum h_\alpha$.

{}From the general theory of differential equations, one would expect
the above formalism to extend to the case where there are logarithmic
solutions. However, unlike ordinary RCFTs, the characters of the
irreducible representations of LCFTs by themselves do not provide
a representation of the modular group\cite{Flohr:1995ea}. The character
of a primary is also the trace (upto a phase) of $q^{L_0}$ over the
module above it. Now, the action of $q^{L_0}$ on the indecomposable
representation in Eq.\refb{jpair} gives us\cite{Flohr:1995ea}
\begin{equation}
q^{L_0}\pmatrix{
\phi_1\cr
\phi_3}
\sim \pmatrix{
q^{L_0^{(1)}} & 0\cr
\ln q\,q^{L_0^{(1)}} & q^{L_0^{(3)}}}\pmatrix{
\phi_1\cr
\phi_3},
\label{qlnot}
\end{equation}
where, the superscripts on $L_0$ refer to its action on the two field
in the Jordan block. In taking a trace to obtain the characters, one
gets two identical series in $q$, equal to the character over the module
over $\phi_1$. Fortunately, there is another set of objects, namely the
one-point correlation functions on the torus, or the {\em vacuum torus
amplitudes} (VTA), which still close under modular transformations. MDEs
were originally written for the VTAs in \cite{Eguchi:1987qd}.

Following \cite{Flohr:1995ea,Flohr:2005cm}, the analysis of Mathur {\em
et al} can be generalised and applied to LCFTs. The solutions of the
MDE are now the vacuum torus amplitudes $T(\tau)$.
In \cite{Mathur:1988na}, it was assumed that the highest weights are all
distinct $h_\alpha\ne h_\beta$ for $\alpha\ne\beta$. Now it is relaxed to
allow for degenerate weights, but the solutions are assumed to
the distinct $T_\alpha\ne T_\beta$ for $\alpha\ne\beta$. The rest of the
analysis is the same as in the case of RCFTs. In particular, LRCFTs can
also be classified by $n$ and $\ell$.

It turns out to be convenient to change variable from $\tau$ to
\begin{equation}
\lambda = \left({\vartheta_2(\tau)\over\vartheta_3(\tau)}\right)^4
= 16 q^{1/2}\left(1 - 8q^{1/2} + 44q + \cdots\right),\label{lq}
\end{equation}
where $q=\exp(2\pi i\tau)$ and $\vartheta$'s are the standard Jacobi
theta-functions\cite{Eguchi:1987qd, Mathur:1988rx}. The above maps
(six copies of) the moduli space to the complex plane which has no
orbifold singularity. In fact, $\lambda$ is the parameter that appears
in the elliptic equation defining the torus. The variable $q$ can in
turn be written as $q = \left({\lambda\over 16}\right)^2\left( 1 +
\lambda + 232 \left({\lambda\over 16}\right)^2 +
\mathcal{O}(\lambda^{3})\right )$.
Using the properties of the theta-functions, one finds that under the
generators $T$ and $S$ of the modular group:
\begin{equation}
\lambda(\tau+1) = {\lambda(\tau)\over\lambda(\tau) -1},\qquad
\lambda\left(-{1\over\tau}\right) = 1- \lambda(\tau).
\end{equation}
Notice that $T^2:\lambda\rightarrow\lambda$, therefore, $\lambda$
really parametrises the subgroup $\Gamma(2)$ of the modular group.
When written in terms of $\lambda$, the coefficients of the MDE are
rational functions and therefore, the MDE is a differential equation
of the Fuchsian type. This is a real advantage of going over to the
variable $\lambda$. For details and additional comments, see
Ref.\cite{Kiritsis:1988kq}.

In the case of the RCFTs, this fact has been used in
\cite{Mathur:1988na}, where theories with two and three characters
were analysed in detail. In the former case, the MDE is a
hypergeometric equation \refb{hyperg}. The authors of
\cite{Mukhi:1989qk} exploit the Fuchsian character of the MDE
written in terms of $\lambda$ to write an explicit set of solutions
as contour integrals --- one does not really need to know the MDE.
The building blocks are the Feigin-Fuchs-Dotsenko-Fateev (DFFF) type
contour integral\cite{Dotsenko:1984ad}: 
\begin{eqnarray}
J &\equiv&  \left(\lambda(1-\lambda)\right)^\alpha
\mathcal{J}(\lambda)\nonumber\\
&\sim& \left(\lambda(1-\lambda)\right)^\alpha \int
\prod_{i=1}^{n_1} dt_i\,  \prod_{k=1}^{n_2} ds_k\,
\prod_{i=1}^{n_1}\left[ t_i (t_i -1) (t_i - \lambda)\right]^a
\prod_{k=1}^{n_2}\left[ s_k (s_k -1) (s_k - \lambda)\right]^b\nonumber\\
& & \qquad\qquad\qquad\prod_{i<j} \left(t_i - t_j\right)^{-2a/b}\;
\prod_{k<l} \left(s_k - s_l\right)^{-2b/a}\; \prod_{i<k} \left(t_i -
s_k\right)^{-2b},\label{ffdf}
\end{eqnarray}
where,
\begin{equation}
\alpha = {1\over {3}} \left( - n_1(1+3a) - n_2(1+3b) + {a\over b}
n_1(n_1 - 1) + {b\over a} n_2(n_2 - 1) + 2 n_1 n_2 \right )
\label{alpha}
\end{equation}
These integrals are invariant under modular $S$ and $T$ transformations
up to changes in integration limits and phase factors. The limits of
the integration are chosen such that the $T$ transformation leaves the
integral invariant upto a phase. These lead to the following set of
$n = (n_1+1)(n_2+1)$ independent integrals $J_{AB}$:
\begin{eqnarray}
J_{AB} &\equiv &  \left(\lambda(1-\lambda)\right)^\alpha
\mathcal{J}_{AB} (\lambda)\nonumber\\
&=&  \left(\lambda(1-\lambda)\right)^\alpha
\prod_{i=1}^A \int_0^\lambda dt_i\,
\prod_{j=A+1}^{n_1} \int_1^\infty dt_i\,
\prod_{k=1}^B \int_0^\lambda ds_k\,
\prod_{l=B+1}^{n_2} \int_1^\infty ds_l\nonumber\\
& & \qquad\qquad\prod_{i=1}^{n_1}\left[ t_i (t_i -1) (t_i - \lambda)
\right]^a\;\prod_{k = 1}^{n_2} \left[ s_k (s_k -1)
(s_k - \lambda) \right]^b \nonumber \\
& & \qquad\qquad\prod_{i<j} \left(t_i - t_j\right)^{-2a/b}\;
\prod_{k<l} \left(s_k - s_l\right)^{-2b/a}\;
\prod_{i<k} \left(t_i - s_k\right)^{-2b},\label{jbasis}
\end{eqnarray}
which go into each other under the modular transformations.

As $\lambda \rightarrow 0$
\begin{equation}
J_{AB}(\lambda) \sim \lambda^{\alpha + \Delta_{AB}},
\label{JABasymp}
\end{equation}
where,
\begin{equation}
\Delta_{AB} = A(1+2a) + B(1+2b) - {a\over b} A(A-1) -
{b\over a} B(B-1) - 2AB.\label{deltaAB}
\end{equation}
Let us observe that $\Delta_{00} = 0$ and
\begin{equation}
\sum_{A=0}^{n_1}\sum_{B=0}^{n_2} \left(\alpha + \Delta_{AB}\right)
= {1\over{6}} n(n-1).\label{sumdelta}
\end{equation}
{}From the behaviour of the character $\chi _\alpha$ associated with
a primary field of weight $h_\alpha$:
\begin{equation}
\chi_\alpha \sim \lambda^{2h_\alpha - {c\over 12}},\label{chiasymp}
\end{equation}
in the $\lambda \rightarrow 0$ limit, one identifies the sets
$\left\{2h_\alpha-c/12\right\}$ and $\left\{\alpha+\Delta_{AB}
\right\}$ and finds that
\begin{equation}
\ell \equiv {nc\over{4}} - 6 \sum h_\alpha + {n(n-1)\over{2}} = 0.
\label{ell}
\end{equation}
Therefore, only the characters of those RCFTs with $\ell=0$ may be
expressed as DFFF integrals.

As shown in \cite{Mukhi:1989qk}, only for RCFTs with at most five
characters, the leading asymptotic behaviour of the DFFF integrals
suffice to prove that they are the characters. When the number of
characters $n\ge 6$, one needs the next to leading order terms in
the power series expansion in $q$ in order to match them with the
expected power series of the characters.

\sectiono{Contour integrals for the `characters' of the $c=-2$
theory}\label{contourcm2} We have mentioned earlier that in the case
of the logarithmic minimal CFTs, the solutions of the modular
differential equation are the vacuum torus amplitudes, rather than
the characters. We would expect that the integral representations,
which provide a set of solutions to the MDE, to represent the vacuum
torus amplitudes. In the following, we propose the set of integrals
for the $(p,1)$ series of logarithmic minimal models ${\cal M}_p$.
Mukhi {\em at al} \cite{Mukhi:1989qk} gave a prescription to
identify the parameters $(A,B,a,b)$ in Eq.\refb{jbasis} to those of
the $(p,q)$ minimal models\cite{Belavin:1984vu}. While this does not
directly apply to the logarithmic family ${\cal M}_p$, one may still
determine these parameters consistently, and indeed, the integrals
are simpler for the LCFTs.

\bigskip

Let us first consider the $c= -2$ model ${\cal M}_2$ that we have
discussed in Sec.\ref{lmm}. The fifth order modular differential
equation for the vacuum torus amplitudes $T$ was
derived\cite{Flohr:2005cm} from the existence of a null vector in
the vacuum module of the chiral $W$-algebra:
\begin{equation}
{\mathcal{D}}_\tau^5 T(\tau) + \sum_{k=0}^4 f_k(\tau)
{\mathcal{D}}_\tau^k T(\tau) = 0.\label{mdeform2}
\end{equation}
{}From the expression \refb{ell}, one easily finds that in this theory
$\ell=0$. Thus the Wronskian does not have a zero for any finite
value of $\tau$, and this model ought to be among those for which the
solutions of the MDE admit an integral representation.

We have $n = 5 = (n_1+1)(n_2+1)$. Therefore, either $n_1$ or $n_2$ must
be zero. This in turn implies that either $A$ or $B$ is zero. Since
there is an invariance under the exchange of the pairs
$(A,a) \leftrightarrow (B,b)$, we may choose $B=0$ without any loss of
generality. Following \cite{Mukhi:1989qk}, our proposal is thus:
\begin{equation}
(A,a)= \left(r-1,-{5/8}\right), \qquad
(B,b)= \left(0,{5/2}\right),
\end{equation}
using which we get the expected values for
$\alpha = {1\over 6}\equiv -{c\over 12}$ and
$\Delta_{AB} = 2h_{r,1}$.

Explicitly, we find the following basis for the
integrals\footnote{In the special case of Eq.\refb{jbasis} for $B=0$
this type of integrals were first considered by Selberg and
generalised by Aomoto\cite{Aomoto:1987}.}:
\begin{eqnarray}
J_{00} &=& \left[\lambda(1-\lambda)\right]^{1/6} \int_1^\infty
dt_1\cdots dt_4\, \prod_{i = 1}^4 \left[ t_i (t_i -1) (t_i - \lambda)
\right]^{-5/8}\,\prod_{i<j}\left(t_i - t_j\right)^{1/2}\nonumber\\
J_{10} &=& \left[\lambda(1-\lambda)\right]^{1/6} \int_0^\lambda
dt_1 \int_1^\infty dt_2 dt_3 dt_4\,\prod_{i = 1}^4 \left[ t_i (t_i -1)
(t_i - \lambda) \right]^{-5/8}\,\prod_{i<j} (t_i - t_j)^{1/2}\nonumber\\
J_{20} &=& \left[\lambda(1-\lambda)\right]^{1/6}  \int_0^\lambda
dt_1 dt_2\int_1^\infty dt_3 dt_4\,\prod_{i = 1}^4 \left[ t_i (t_i -1)
(t_i - \lambda) \right]^{-5/8}\,\prod_{i<j} (t_i - t_j)^{1/2}
\label{jbasism2}\\
J_{30} &=& \left[\lambda(1-\lambda)\right]^{1/6} \int_0^\lambda
dt_1 dt_2 dt_3 \int_1^\infty dt_4\,\prod_{i = 1}^4 \left[ t_i (t_i -1)
(t_i - \lambda) \right]^{-5/8}\,\prod_{i< j} (t_i - t_j)^{1/2}\nonumber\\
J_{40} &=& \left[\lambda(1-\lambda)\right]^{1/6} \int_0^\lambda
dt_1\cdots dt_4 \prod_{i = 1}^4 \left[ t_i (t_i -1) (t_i - \lambda)
\right]^{-5/8}\prod_{i<j} (t_i - t_j)^{1/2}\nonumber
\end{eqnarray}
Since, this is a theory with five `characters', it will suffice to
match the leading asymptotic behaviour in order to identify the DFFF
integrals with the vacuum torus amplitude. The leading $\lambda$
dependence suggests that $J_{A0}$ corresponds to the vacuum torus
amplitude of the field $\phi_{A+1}$. However, both $J_{00}$ and
$J_{20}$ correspond to fields with $h=0$ and the weight of the field
for $J_{40}$, $h=1$, differs from these by an integer. So a change
of basis may be necessary. Moreover, the set above cannot be
linearly independent and we will need to put in the logarithmic
solution corresponding to the degenerate roots $h=0$ of
\refb{mdeform2}.

\bigskip

Let us now compute the subleading terms. We consider $J_{00}$ first.
Making a change of variable to bring the limits of the integrals in
standard form, we rewrite it as (we have not put the specific values
of $\alpha, a$ and $b$ at this stage):
\begin{equation}
J_{00} = \lambda^\alpha\int_0^1\prod_{i=1}^{n_1}dt_i\,
t_i^{-2-3a+{2a\over b}(n_1 -1)} (1-t_i)^a
\prod_{i<j} (t_i -t_j)^{-{2a\over b}}\, F(\lambda),\label{jasf}
\end{equation}
where,
\begin{eqnarray}
F(\lambda) &=& (1- \lambda)^{\alpha}\, \prod_{i=1}^{n_1}
(1-t_i \lambda)^a
\;=\; 1 - \left( \alpha + a \sum_{i=1}^{n_1} t_i\right)\lambda
\label{flambda}\\
&&+ \left( {\alpha(\alpha -1)\over{2}} + a \alpha \sum_{i=1}^{n_1}
t_i + {a(a -1)\over2} \sum_{i=1}^{n_1} t_i^2 + a^2 \sum_{i<j}t_i t_j
\right) \lambda^2 + \mathcal{O}(\lambda ^3)\nonumber
\end{eqnarray}
The integrals that one needs to evaluate at each order in $\lambda$,
when the expansion \refb{flambda} is substituted in Eq.\refb{jasf},
have fortunately been evaluated in \cite{Dotsenko:1984ad}. Following
the notation there, let us define:
\begin{equation}
\left\{ f(t_i)\right\} = \prod _{i=1}^{n}\int_0^1 dt_i\, t_i^\gamma
(1-t_i)^\delta \prod_{i<j} (t_i - t_j)^{2\rho} f(t_i),
\label{dotfatint}
\end{equation}
where $\gamma, \delta, \rho$ are arbitrary constants and $f(t_i)$
is a function of $t_i$.

The expansion of $J_{00}$ in $\lambda$ is therefore:
\begin{eqnarray}
J_{00}(\lambda) &=& \lambda^\alpha \Bigg[ \{1 \} - \left(\alpha \{1\}
+ n_1 a \{t_1\}\right)\,\lambda + \Bigg({\alpha(\alpha -1)\over{2}}
\{1\}+ a \alpha n_1\{t_1\} \nonumber \\
&& \qquad + {a(a -1)\over{2}}\,n_1 \{t_1^2\} + a^2 {n_1(n_1 -1)\over
2} \{t_1 t_2\} \Bigg)\lambda^2 +\mathcal{O}(\lambda ^3) \Bigg],
\label{j00expansion}
\end{eqnarray}
where $\{1\}$ is the overall normalisation constant which we may
choose to factor out and define the normalised integrals
$\overline{f(t_i)} = { \{ f(t_i) \} / { \{ 1\}}}$.

The integrals \refb{dotfatint} are special cases of the more general
DFFF integrals. The case $f(\{t_i\})=1$, {\em i.e.}, the
normalisation constant, was evaluated by Selberg:
\begin{equation}
\left\{1\right\} =
\prod_{m=0}^{n-1}{\Gamma\left((m+1)\rho\right)\Gamma(1+\gamma+m\rho)
\Gamma(1+\delta+m\rho)\over\Gamma(\rho)
\Gamma(2+\gamma+\delta+(n-1+m)\rho)}.\label{one}
\end{equation}
This result was generalised by Aomoto \cite{Aomoto:1987} to the {\em
elementary symmetric polynomials}
\begin{equation}
P^{(n)}_k(\{t_i\}) = \displaystyle\sum_{1\le j_1<\cdots<j_k\le n}
\displaystyle\prod_{i=1}^k t_{j_i} = {1\over \Gamma(k+1)
\Gamma(n-k+1)}\sum_{\sigma_n}\prod_{i=1}^k\,t_{\sigma_n(i)}\label{esp}
\end{equation}
of degree $k$ and found to be:
\begin{equation}
\overline{P^{(n)}_k(\{t_i\})} = {\Gamma(n+1)\over\Gamma(k+1)
\Gamma(n-k+1)}\, {\Gamma\left({\gamma+1\over\rho}+n\right)
\Gamma\left({\gamma+\delta+2\over\rho}+2n-k-1\right) \over
\Gamma\left({\gamma+1\over\rho}+n-k\right)
\Gamma\left({\gamma+\delta+2\over\rho}+2n-1\right)}.\label{aomotoint}
\end{equation}
Let us list the ones we need for calculation up to order $\lambda^2$
for completeness: From the above we find:
\begin{eqnarray}
\overline{t_1} \,=\, {1\over n}\overline{P^{(n)}_1}\; &=& {\gamma +
1 + (n -1) \rho \over \gamma + \delta + 2 + 2(n -1) \rho},\nonumber\\
\overline{t_1 t_2} \,=\, {2\over n(n-1)}\overline{P^{(n)}_2}\;&=&
{\left(\gamma + 1 + (n -1) \rho\right)\left( \gamma + 1 + (n -2)
\rho\right) \over \left(\gamma + \delta +2+ 2 (n -1)
\rho\right)\left(\gamma + \delta +2+ (2n -3) \rho\right)},
\label{FFint}
\end{eqnarray}
and in addition, the following integral has been evaluated in
\cite{Dotsenko:1984ad}:
\begin{equation}
\overline{t_1^2} = \overline {t_1 t_2} + {\left(1+n \rho\right)
\left(\gamma + 1 + (n -1) \rho\right)\left(\delta + 1 + (n -1) \rho
\right) \over \left(\gamma + \delta +2+ 2(n -1) \rho\right)\left(
\gamma + \delta +2+ (2n -3) \rho\right)\left(\gamma + \delta +3+ 2(n
-1) \rho\right)}.\label{FFinttwo}
\end{equation}
The relevant constants in this case are $\gamma = -13/8$, $\delta =
-5/8$ and $\rho= 1/4$: hence $\overline {t_1} = {1/{10}}$,
$\overline {t_1 t_2} = -{1/{80}}$ and $\overline{t_1^2}= {7/{80}}$.
Let us return to the normalisation constant, the exact value of
which is not important at this point, later. Suffice to say that it
is a finite constant. This brings us to the expansion (determined up
to an overall normalisation constant): 
\begin{eqnarray}
J_{00} &\sim & \lambda^{1/6} \left( 1 + {1\over{12}}\lambda
+ {43\over{1152}}\, {\lambda}^2 + \mathcal{O}(\lambda^3) \right),
\nonumber\\
&\sim & q ^{1/12} \left( 1 + \mathcal{O} (q^2) \right ).
\label{idchar}
\end{eqnarray}
where we have used Eq.\refb{lq} in writing the last expression in
terms of $q$.

The first subleading term in the $q$ expansion is thus zero, which
implies the existence of a null vector at the first level. This is
expected of the identity field. Since the identity field is unique
in a CFT, we identify $J_{00}$ to the one-point function of the
identity field $\phi_1$ at one loop.

It is straightforward to repeat the steps above to find the
subleading terms in the other integrals. We find (see
Eq.\refb{JAzero} to order $\lambda^2$: 
\begin{eqnarray}
J_{10} &\sim & q^{-1/24}\left(1+q+\mathcal{O}(q^2)\right),\nonumber\\
J_{30} &\sim & q^{11/24} \left( 1 + q + \mathcal{O} (q^2) \right),
\label{jexpandq}\\
J_{40} &\sim & q ^{13/12}\left(1+q+\mathcal{O}(q^2)\right),\nonumber
\end{eqnarray}
up to the overall normalisation constants which we can determine
(see Eq.\refb{normalisation}). The fact that the coefficient of the
term $q^{1/2}$ vanishes in all cases provides a non-trivial check.

\bigskip

So far we have not discussed the integral $J_{20}$. This is supposed
to describe the VTA of the partner of the identity field in the
Jordan block and exhibit logarithmic behaviour. The normalisation
plays an important part here. The two sets of variables for the two
integrals require different redefinitions to bring them to the
standard form in Eqs.\refb{dotfatint} (see Eq.\refb{JAzero}). The
normalisation for the second set has a factor of
$\Gamma(2+\gamma+\delta+\rho)$ corresponding to $m=0$ in
Eq.\refb{one}. The argument for the values of the parameters in the
second integral vanishes, leading to a divergent factor of
$\Gamma(0)$ in the denominator. This would make $J_{20}$ in
\refb{jbasism2} vanish. Interestingly, this is not quite the case.
Even though the (normalised) coefficients of the terms of order one
and $\lambda$ are finite, the same at order $\lambda^2$ diverges.
Moreover, the divergence is due to a vanishing factor in the
denominator that is exactly the argument of the gamma function
mentioned above. This renders it finite. Put in a different way, the
divergence is an artefact of separating out the normalisation
factor. The expansion of the integral $J_{20}$ as a power series in
$\lambda$ actually starts at order $\lambda^2$. The same conspiracy
of factors give finite coefficients for all the higher order terms
(since the same factors are part of the integrals \refb{aomotoint}
that appear at subsequent orders). Thus we find an infinite series
in $\lambda$, the leading power of $\lambda$ (and hence $q$) of
which cannot correspond to the field $\phi_3$ with $h_3=0$. Indeed,
since there is only one root of the indicial equation with this
power, $J_{20}$ cannot be an independent solution of the MDE, rather
it must be the same as $J_{40}$ up to an overall constant. Moreover,
this being a model with five fields, the leading order behaviour
suffices to identify the corresponding character uniquely. In
principle, one should also be able to check this explicitly, at
least term by term in the power series. However, in practice, this
requires the values of the integrals \refb{dotfatint} with
polynomials of increasing order. Only a subset of these integrals
corresponding to the elementary symmetric polynomials \refb{esp}
have been evaluated\cite{Aomoto:1987}. One can check from
Eq.\refb{aomotoint} that the series from $J_{20}$ is of the expected
form.

We, therefore, need to substitute $J_{20}$ by a linearly independent
solution. To this end, we need a regularisation scheme to replace
the vanishing factor by $\varepsilon$ and in addition we propose
$\alpha\rightarrow\alpha+\eta$. One may think of the latter as an
analytic continuation in the central charge. As a result, we get
$J_{A0}(\lambda,\varepsilon,\eta)$ as a series in $\lambda$,
$\varepsilon$ and $\eta$. We find that if we set $\varepsilon=\eta$
and consider the linear combination 
\begin{equation}
\tilde{J}_{20} \equiv \lim_{\varepsilon\to 0} \left[ {1\over
\varepsilon^2}\,J_{20} + \left( a_0 + {a_1\over
\varepsilon}\right)\,J_{00} + \left(b_0 + {b_1\over\varepsilon} +
{b_2\over\varepsilon^2}\right)\,J_{40} \right],\label{logsoln}
\end{equation}
where, $a_0,\cdots,b_1$ are appropriately defined constants, the
unwanted divergent terms cancel out. The resultant finite terms
behave like $\ln\lambda$ times an infinite series in $\lambda$ with
the leading power of order one. This is the linearly independent
logarithmic solution of the MDE.

\bigskip

We are now in a position to compare with the results of
Ref.\cite{Flohr:1995ea,Flohr:2005cm}, where the authors constructs the
characters of the $c=-2$ triplet algebra. These are expressed in terms
of the Dedekind eta- and generalized theta-functions:
\begin{eqnarray}
\eta(q) &=& q^{1/24} \prod _{n=1}^\infty (1-q^n),\nonumber\\
\theta_{\nu,k}(q) &=& \sum_{n=-\infty}^\infty q^{(2kn+\nu)^2/4k},
\label{gentheta}\\
{\partial \theta}_{\nu,k}(q) &=& \sum_{n=-\infty}^\infty (2kn+\nu)
q^{(2kn+\nu)^2/4k}.\nonumber
\end{eqnarray}
The characters of the four irreducible highest weight representations
of the chiral algebra, namely, the identity ($h=0)$ and the three
fields with conformal weights $h=-1/8, 3/8$ and $1$, are:
\begin{eqnarray}
\chi_{0}(q) &=& \left(\theta_{1,2}+{\partial\theta}_{1,2}
\right)/\eta\;=\; 2 q^{1/12}\left(1+q^2+4q^3\cdots\right),\nonumber\\
\chi_{-1/8}(q) &=& \theta_{0,2}/\eta\;
=\; q^{-1/24}\left(1+q+4q^2+5q^3\cdots\right),\nonumber\\
\chi_{3/8}(q) &=& \theta_{2,2}/\eta\;=\; 2q^{11/24} \left(1+q+
2q^2+3q^3+\cdots\right),\nonumber\\
\chi_{1}(q) &=& \left(\theta_{1,2}-\partial\theta_{1,2}\right)/
\eta\;=\; 4 q^{13/12}\left(1+q+q^2+2q^3+\cdots\right).
\label{flohrchi}
\end{eqnarray}
Using these results, the corresponding vacuum torus amplitudes have
been proposed in \cite{Flohr:2005cm} by analysing the modular
differential equation order by order in $q$. One may take the VTAs
of the identity and the fields $\phi_2$, $\phi_4$ and $\phi_5$ to be
the same as the respective characters above. In addition, the linearly
independent solution
\begin{equation}
T(q) = \ln q\,\partial\theta_{1,2}/\eta\label{logchi}
\end{equation}
may be taken to be the VTA of the field $\phi_3$. Of course, as has
been stressed in \cite{Flohr:1995ea,Flohr:2005cm}, there is no
canonical choice of basis in this case and the proposal above is one
possible consistent choice.

A comparison with \refb{flohrchi} justifies our proposal for the
vacuum torus amplitudes to the DFFF integrals: $J_{00}$ to the
identity, $J_{10}$ to $\phi_2$, $J_{30}$ to $\phi_4$ and $J_{40}$ to
$\phi_5$. In addition, in both cases, we have a solution with
logarithmic dependence as the vacuum torus amplitude of the field
$\phi_3$. In \cite{Flohr:2005cm}, Eq.\refb{logchi} is obtained by
subtracting $\chi_0$ and $\chi_1$ and multiplying by a $\ln q$
piece. If we substitute the integral representation that we have
already identified, we find:
\begin{equation}
\ln q\left(\chi_0-\chi_1\right) = \ln
q(\lambda)\,\left[\lambda(1-\lambda)\right]^{{1\over6}}
\int_{1}^{\infty}\!\prod_{i = 1}^4 dt_i \left( t_i (t_i-1)
(t_i-\lambda) \right)^{-{5\over8}}\left(1-{\lambda^2\over
t_i}\right) \prod_{i<j} (t_i - t_j)^{{1\over2}}. \label{logff}
\end{equation}
This is evidently of the form that one obtains from the combination
\refb{logsoln}. The new set of integrals solves the modular
differential equation and is closed under modular transformations.

\sectiono{Contour integrals for the `characters' of the $c_{p,1}$ models}
\label{contourlmm}

The contour integral representation for the vacuum torus amplitudes
of the $c=-2$ logarithmic conformal field theory generalises to the
other logarithmic minimal models ${\cal M}_{p\ge3}$ of the $(p,1)$
family. In this section, we will briefly indicate how to extend our
proposal. First, we compute that the value of $\ell$ for any model
of the series and find that:
\begin{equation}
\ell = {(3p -1)(3p-2)\over2} + {3p-1\over4}\left(1 - {6(p-1)^2\over{p}}
\right) - 6 \sum _{r=1}^{3p-1} {{(p-r)^2 - (p-1)^2}\over{4p}} = 0.
\nonumber
\end{equation}
So, there is no obstruction for an integral representations for the
vacuum torus amplitudes for these theories. Next we need to identify the
set $(A, B, a, b)$ with the parameters of ${\cal M}_p$. The order of the
MDE for the vacuum torus amplitudes is $3p -1$, and hence,
$(n_1+1)(n_2+1) = 3p -1$. Unlike the case of the $c=-2$ model, a priori
there no reason now for $B$ to be zero. Nevertheless, unless $B=0$, we
are led to a contradiction.

To see this, let us recall that the set $\Delta_{AB}$ is to be identified
with $2 h_{r,1}$. Since only $A$ and $B$ can depend on $r$, this leads to
the relation
\begin{equation}
A + 2pB = r-1, \label{ineq}
\end{equation}
after we have identified
\begin{equation}
a = {{3-4p}\over{4p}}, \qquad b = -{{3-4p}\over{2}},\label{abcp1}
\end{equation}
following \cite{Mukhi:1989qk}.
The fact that $A=0$ for the maximum value of $B$ and $r\le 3p-1$, gives
the bound $2pB \le 3p -2$. The only allowed integer values are therefore
$B=0,1$ for any $p\ge 2$. If $B=1$, we get $n_2=1$. Substituting this
into the relation $\alpha = -{c\over 12}$ and using \refb{alpha} we get
a quadratic equation for $n_1$, the solutions of which are
$n_1=5p-2$ and $n_1=3p-{3\over2}$. On the other hand, from the condition
$(n_1+1)(n_2+1) = 3p -1$, we find $n_1 = {3\over 2} (p-1)$.
Consistency of these require $p= 1/7$ or $p=0$, neither of which is
acceptable. Hence $B=n_2=0$ for all the $(p,1)$ models.

The following relation between the two sets of parameters:
\begin{equation}
(A,a,n_1) = \left(r-1, {3-4p\over{4p}},3p -2\right),\qquad
(B,b,n_2) = \left(0, {{4p-3}\over{2}}, 0\right),\label{paramcp1}
\end{equation}
gives rise to correct values for $\Delta_{AB}={(p-r)^2-(p-1)^2\over
2p}$ and $\alpha = -{c\over {12}}$. A basis for the DFFF integrals
is \refb{jbasis} with the values of the parameters as above.
However, this set cannot be linearly independent.

We see from the conformal weights \refb{cweights}, that the pairs of
integrals $J_{A0}$ and $J_{A'0}$ with $A+A'=2p-2$,
$A=0,2,\cdots,(p-2)$ have the same leading power of $\lambda$. This
is consistent with the Jordan block structure of these models
discussed in Sec.\ref{lmm}. Moreover, the leading powers of
$\lambda$ of $J_{A0}$ for $A=2p,\cdots,3p-2$ differ from the former
sets respectively by $2r=2,4,\cdots,2p-2$, all even integers. This
suggests linear relations involving the triads of integrals
$\left(J_{r-1,0},J_{2p-r-1,0},J_{2p+r-1,0}\right)$. The field
$\phi_r$ is an eigenfunction of $L_0$ and we expect $J_{r-1,0}$ to
be its VTA.

Let us examine the behaviour of $J_{A0}$. There are two sets of
integrals with different limits. After changing variables to bring
them to the standard form we find:
\begin{eqnarray}
J_{A0} &=& \lambda^{\alpha + A(1+2a-a(A-1)/b)}\, \prod_{i=1}^A
\int_0^1 ds_i\,
\left(s_i(s_i-1)\right)^a \prod_{i<j}(s_i-s_j)^{-2a/b}\nonumber\\
&&\times\,\prod_{k=A+1}^{n_1}\int_0^1 du_k\,
u_k^{-2-3a+2a(n_1-1)/b}(1-u_k)^a \prod_{k<l}(u_k-u_l)^{-2a/b}
\label{JAzero}\\
&&\times\, \left(1-\lambda\right)^\alpha\,\prod_{i=1}^A (1-\lambda
s_i)^a\,\prod_{k=A+1}^{n_1} (1-\lambda u_k)^a
\,\prod_{i,k}(1-\lambda s_i u_k)^{-2a/b},\nonumber
\end{eqnarray}
which may be expanded to the desired order in $\lambda$.  Consider
the overall normalisation factor in Eq.\refb{JAzero}:
\begin{equation}
{\cal N}_{A0}\,=\, \prod_{m=0}^{A-1}{\Gamma\left({m+1\over
2p}\right)\left(\Gamma\left({2m+3\over 4p}\right)\right)^2\over
\Gamma\left({1\over 2p}\right)\Gamma\left({m+A+2\over 2p}\right)}\;
\prod_{n=0}^{3p-3-A}{\Gamma\left({n+1\over 2p}\right)
\Gamma\left({2n+3\over 4p}-1\right) \Gamma\left({2n+3\over4p}
\right)\over\Gamma\left({1\over 2p}\right)\Gamma\left({p-A+n\over
2p}\right)}. \label{normalisation}
\end{equation}
The normalisation constant is finite for the integrals $J_{A0}$ for
$A=0,1,\cdots,p-2$, as well as $A=p-1$ and $A=2p-1,\cdots,3p-2$.
However, rather interestingly one finds that for $A=p,\cdots,2p-2$,
all these, and only these, integrals have a divergent factor of
$\Gamma(0)$ in the denominator from $n=0,1,\cdots,p-2$ in
Eq.\refb{normalisation} respectively! Recall that these were to
correspond to the VTA for the Jordan block partner of the first set
(in reverse order).

As in the $c=-2$ case discussed earlier, this would have made the
integrals vanish. However, just like in that case, a zero appears in
the denominator of the coefficient of a higher order term in
$\lambda$, so as to give a finite expression! Indeed, the
coefficient of term at order $\lambda^2$ in the expansion of the
integral in $J_{2p-2,0}$ does diverge and this makes it finite.
Similarly in $J_{2p-2-A,0}$ ($A=0,\cdots,p$), one can check that
precisely the coefficient of the term $\lambda^{2A+2}$ (we are
counting modulo the power due to the weight of the field) in
\refb{aomotoint} has a divergent factor from $k=2A+2$  that makes it
finite. The same factor is part of the integrals that appear in
subsequent orders. Thus the behaviour of these integrals is not
right to describe the VTAs of the fields $\phi_{p+1},\cdots,
\phi_{2p-1}$, but rather they ought to be equal to $J_{2p+A,0}$
($A=0,\cdots,p$). Therefore, as in the $c=-2$ model, we propose to
regularise the triad ($J_{A0}$, $J_{2p-2-A,0}$, $J_{2p+A,0}$) and
replace the redundant second integral by a linear combination (see
Eq.\refb{logsoln}) which exhibits logarithmic dependence in
$\lambda$ and hence in $q$. This prescription provides the required
`logarithmic characters'.

\bigskip

In the general case, there is another issue to resolve. Just the
leading order behaviour is not enough to establish that these are
indeed the vacuum torus amplitudes. Since the number of `characters'
$3p-1>5$ for $p>2$, one also needs to analyse the nonleading terms
in this case. This could be done in a fashion similar to the $c=-2$
case. Let us consider the integral $J_{00}$ which ought to
correspond to the VTA of the identity field. Using
Eqs.\refb{jasf}--\refb{FFinttwo}, we find
\begin{equation}
J_{00} \sim \lambda^\alpha\left(1 + {1\over 2}\alpha\,\lambda +
{1\over 8}\alpha\left(\alpha + {13\over 8}\right)\lambda^2 +
\mathcal{O}(\lambda^3)\right),\label{idp1}
\end{equation}
where $\alpha=-{c\over 12}={1\over 2}\left(p + {1\over p} - {13
\over 6}\right)$. Hence, using Eq.\refb{lq}, we find that the
coefficients of the $q^{1/2}$ and $q$ terms vanish, so that
$J_{00}\sim q^{-c/24}\left(1+\mathcal{O}(q^2)\right)$ as expected of
the identity field. This can be repeated for the other integrals.

\sectiono {Conclusions}\label{concl}
We have proposed a contour integral representation for the vacuum
torus amplitudes (one point function on the torus) of the
logarithmic $(p,1)$ minimal models, generalising earlier results for
the characters of ordinary rational conformal field theories. We
have studied the simplest $(2,1)$ model with central charge $c=-2$
in some detail and identified the candidate integrals for the other
models of the family. In fact, in the case of these logarithmic
family of $(p,1)$ minimal models, the integral representation (being
of the Selberg-Aomoto type, rather than the more general type
considered by Feigin-Fuchs and Dotsenko-Fateev) is simpler than the
well known RCFT case.

For the irreducible representations, as well as for the fields in
the indecomposable Jordan block that are the eigenfunctions of
$L_0$, we find the integral representations by extending the known
results of the RCFTs. The same procedure applied to the partner
fields in the Jordan block may seem to naively yield a vanishing
expression. These are, however, the VTAs of primaries, the weights
of which differ from that of the Jordan cell by an integer. We
propose to replace these by the corresponding logarithmic characters
obtained by a regularisation prescription. The regularisation
scheme, however, is chosen for simplicity and could perhaps be
improved upon for better understanding.

The modular differential equation for the $c=-2$ model was analysed
as a power series in $q$ in \cite{Flohr:2005cm}. It may be
interesting to study the MDE in $\lambda$, as this is a Fuchsian
equation. In principle, one can solve for the recursion relations
and obtain the series solutions. This approach, as well as the
integral representation to generate the series could be useful in
view of the fact that the solutions of the MDEs for the VTAs are not
always expressible in terms of known transcendental
functions\cite{Gaberdiel:2008ma}.


Let us close with a few brief remarks. First, there is a well
defined relation between the minimal models RCFTs and the affine
SU(2) CFTs. This helps to relate the characters of the two theories.
The logarithmic extension of $\mbox{\rm SU(2)}_k$ theories have been
studied by Nichols\cite{Nichols:2002qv}, who shows that the
Hamiltonian reduction of the $\mbox{\rm SU(2)}_{k=0}$ theory yields
the logarithmic $c=-2$ model. The contour integral representation
may be useful in extending this connection to the entire logarithmic
family. It is interesting to note that the contour integrals for the
characters of the ordinary $\mbox{\rm SU(2)}_k$ models are also
given by Selberg type integrals ($B=0$). Secondly, the modular
differential approach to the characters have been extended to the
genus two case in Ref.\cite{Mathur:1988xc}, where the MDE is written
for one of the three complex moduli (keeping the others fixed).
Since genus two surfaces are hyperelliptic, in terms of their
representation as a sphere with six branch points, the MDE is
Fuchsian. A contour integral representation for the genus two
characters should therefore be possible. Finally, it may be an
interesting problem to extend this approach to the various other
known logarithmic conformal field theories.

\bigskip\bigskip

\noindent{\bf Acknowledgments:} It is a pleasure to thank Shamik
Banerjee, Mathias Gaberdiel, Dileep Jatkar, Sunil Mukhi, Sudhakar
Panda, Rajesh Gopakumar for useful discussions and especially Ashoke
Sen for many valuable suggestions as well as critical comments on
the manuscript. TB and DG would like to thank the organisers and
participants of the international workshop on Logarithmic CFTs and
Statistical Mechanics, Dubna, Russia for an opportunity to learn the
subject. AB and TB acknowledge the School of Physical Sciences,
Jawaharlal Nehru University, for hospitality during a part of this
work.

\newpage



\end{document}